# Urban greenery and mental wellbeing in adults: Cross-sectional mediation analyses on multiple pathways across different greenery measures

Ruoyu Wang[a,b,*], Marco Helbich[c,*], Yao Yao[d], Jinbao Zhang[a,b], Penghua Liu[a,b], Yuan Yuan[a,b,#], Ye Liu[a,b,#]

* These authors contributed equally

[a] School of Geography and Planning, Sun Yat-sen University, Guangzhou, 510275, China; email: liuye25@mail.sysu.edu.cn (Y. Liu); kampau@foxmail.com (J. Zhang); liuph3@mail2.sysu.edu.cn (P. Liu); yyuanah@163.com (Y. Yuan)

[b] Guangdong Key Laboratory for Urbanization and Geo-Simulation, Sun Yat-sen University, Guangzhou, 510275, China

[c] Department of Human Geography and Spatial Planning, Utrecht University, The Netherlands; email: m.helbich@uu.nl

[d] School of Information Engineering, China University of Geosciences, Wuhan, 430074, China; email: yaoy@cug.edu.cn

Corresponding authors:

Yuan Yuan and Ye Liu, Sun Yat-sen University, 135 Xingang Xi Road, Guangzhou, 510275, China; tel.: +86 136 024 612 81; email: yyuanah@163.com and liuye25@mail.sysu.edu.cn

*Abstract*

*Background:* Multiple mechanisms have been proposed to explain how greenery in the vicinity of people's homes enhances their mental health and wellbeing. Mediation studies, however, focus on a limited number of mechanisms and rely on remotely sensed greenery measures, which do not accurately capture how neighborhood greenery is perceived on the ground.

*Objective:* To examine: 1) how streetscape and remote sensing-based greenery affect people's mental wellbeing; 2) whether and, if so, to what extent the associations are mediated by physical activity, stress, air quality and noise, and social cohesion; and 3) whether differences in the mediation across the streetscape greenery and NDVI exposure metrics occurred.

*Methods:* We used a population sample from 2016 of 1,029 adult residents of the metropolis of Guangzhou, China. Mental wellbeing was quantified by the World Health Organization Well-Being Index (WHO-5). Two objective greenery measures were extracted at the neighborhood level: 1) streetscape greenery from street view data via a convolutional neural network, and 2) the normalized difference vegetation index (NDVI) from Landsat 8 remote sensing images. Single and multiple mediation analyses with multilevel regressions were conducted.

*Results:* Streetscape and NDVI greenery were weakly and positively, but not significantly, correlated. Our regression results revealed that streetscape greenery and NDVI were, individually and jointly, positively associated with mental wellbeing. Significant partial mediators for the streetscape greenery were physical activity, stress, air quality and noise, and social cohesion; together, they explained 62% of the association. For NDVI, only physical activity and social cohesion were significant partial mediators, accounting for 22% of the association.

*Conclusions:* Mental health and wellbeing and both streetscape and satellite-derived greenery seem to be both directly correlated and indirectly mediated. Our findings signify that both greenery measures capture different aspects of natural environments and may contribute to people's wellbeing by means of different mechanisms.

**Keywords**

Green space; street view data; deep learning; mediation; mental wellbeing; China

**Highlights**

- Mechanisms underlying mental health and wellbeing benefits of greenery are not well established.
- Streetscape greenery from street view images and a remote sensing-based vegetation index (NDVI) served as exposure measure.
- Higher mental wellbeing was correlated with more greenery, independently of the measure.
- Substantial proportions of greenery–wellbeing associations were explained by mediators; for streetscape greenery, mediation was higher.
- Different pathways seem to be at play across both greenery measures.

# 1 Introduction

The support of people's mental health and wellbeing[1] through natural environments is receiving increasing attention [1,2]. Accumulated findings suggest that exposure to outdoor greenery in the vicinity of people's homes is intertwined with and plays a vital role in their mental health and wellbeing [3]. Various cross-sectional and a few longitudinal studies have reported that exposure to pronounced quantities of greenery contributes to mental health and wellbeing [4–9], although this has not always been consistently confirmed [10–13].

However, in what ways exposure to greenery affects mental health and wellbeing remains an open question grounded in a limited empirical evidence base [14,15]. Without reaching an ultimate consensus, the following biopsychosocial pathways were suggested to underlie the health benefits of greenery: Exposure to greenery supports restoration from stress and mental fatigue [16,17], encourages people to be physically active [18–20], filters health-threatening pollutants such as particulate matter and noise [21–23], and enhances neighborhood social cohesion among residents [24,25]. However, while theoretically sound, empirical studies examining one or more of these pathways are significantly underrepresented and often produce contradictory findings [19,24,26,27].

This lack of consensus on the pathways underlying the greenery–mental health relation may also be due to, for example, the way the exposure to greenery is assessed. Greenery is typically represented through remote sensing data [15,28], such as the normalized difference vegetation index (NDVI) [29,30]. It is, however, questionable whether such overhead-view measures from a bird's eye view accurately capture how people on the ground perceive vegetation [31–34]. Executing manual in situ greenery audits, or doing them online through street view services, is a resource-intensive alternative [35,36]. Street view services (e.g., Tencent Online Map, Google Street View) display streetscape images along streets through the web. Recent progress in deep learning [37,38], however, enables the automatic extraction of streetscape greenery, such as trees and green walls, from street view images obtained from online street view services. A preliminary study carried out in Beijing, China, showed that streetscape greenery derived from street view data was inversely correlated with depressive symptoms among elderly people, while this was not the case for NDVI [31]. This suggests that streetscape greenery is particularly vital in dense Chinese inner cities, where larger areas of greenery are often absent. Vegetation reduces the levels of air pollutants and noise from urban traffic [14] and seems to support people's mental restoration. Therefore, we speculated that streetscape and remotely sensed greenery exposure metrics indeed capture diverse aspects of greenery and differ in their underlying pathways (i.e., the number and magnitude of mediators).

---

[1] As a multi-dimensional construct, mental wellbeing constitutes positive states of psychological, life satisfaction, as well as emotional health [63].

In light of these research gaps, we set out to investigate, using a population sample in Guangzhou, China: 1) How streetscape greenery and NDVI are correlated with mental wellbeing; 2) in response to a recent call [14], whether and, if so, to what extent the associations are mediated by physical activity, stress, air quality and noise, and social cohesion; and 3) whether differences in the mediation across the streetscape greenery and NDVI exposure metrics occurred, which if verified, would point toward diverse underlying mechanisms.

## 2 Materials and methods

### 2.1 Study population

This was an observational study conducted in Guangzhou, a city on the Pearl River in mainland China. It is part of one of the world's most urbanized metropolitan areas and has a population of more than 11 million [39]. Examining this metropolis balances the excess of study sites located in the Western world [15].

Data were collected by means of a cross-sectional survey based on face-to-face interviews at people's homes in June–August 2016. To be involved in the survey, respondents had to be aged between 20 and 76 years and have been registered as inhabitants of Guangzhou for at least 12 months. 1,050 respondents were addressed. Respondents were selected through a strict sampling procedure. In brief, a multi-stage stratified probability proportionate to population size sampling design was implemented. This resulted in 35 randomly sampled neighborhoods with a mean area of 1.91 km$^2$ (SD±574.691m$^2$), comprising, on average, 4,155 people (SD±606). The neighborhood level is the most detailed administrative level available in China. The total sample size was 1,029 respondents. The response rate was 98%.

We assessed greenery at the neighborhood level where people live in order to comply with privacy regulations. Using the exact residential address was not possible because it would facilitate to re-identify a respondent. The study received approval from the Sun Yat-sen University Research Ethics Committee and consent was obtained from all subjects.

*2.2 Datasets*

*Mental wellbeing*

The outcome measure was respondents' mental wellbeing. To quantify a multi-dimensional construct like subjective psychological wellbeing, we applied the World Health Organization Well-Being Index (WHO-5) [40], which had previously been implemented in greenery studies [8,13]. The questionnaire comprises five items evaluating respondents' mental health-related feelings over the previous two weeks. Each positively phrased item was scored on a Likert scale with six categories ranging from "at no time" to "all of the time." A Cronbach's alpha of 0.815 referred to an excellent internal consistency for the five items. The individual items were used to receive a single WHO-5 index score ranging from 0 (= worst wellbeing) to 25 (= best wellbeing). Evaluation studies have certified the good validity and reliability of the WHO-5 for general populations [40,41]. A Cronbach's alpha of 0.815 shows that our sample had an excellent internal consistency.

*Greenery data*

Neighborhood-based outdoor greenery was determined in two ways. First, to assess streetscape greenery, we used a series of street view images from Tencent Online Map, which is the most comprehensive street view image database in China [42]. Image locations, each 100 meters apart, were regularly sampled along the road network. Road data for 2016 were obtained from OpenStreetMap [43]. For each sampling location, we downloaded street view images with a dimension of 480×320 pixels from four headings (i.e., 0, 90, 180, and 270 degrees) [44]. This resulted in 31,414 sampling points and 125,656 street view images. On average, we considered 2,106 images per neighborhood (SD=768.016).

To extract streetscape greenery objects (e.g., plants and trees), a supervised machine learning approach based on semantic image segmentation techniques was implemented [31]. More precisely, we used a fully convolutional neural network for semantic image segmentation (FCN-8s) [38] together with the online ADE20K dataset of annotated images for training purposes [45]. For methodological in-depth discussions, we refer to the original references. Based on the pixel-by/to-pixel comparison with a manually conduced segmentation [46], the accuracy of the FCN-8s was reasonably high, namely 0.814 for the training data and 0.800 for the test data. Streetscape greenery per sampling point was determined as the ratio of the number of greenery pixels per image summed over the four cardinal directions to the total number of pixels per image summed over the four cardinal directions. To receive a neighborhood-based streetscape greenery measure, we averaged the image-specific greenery scores per area.

Second, we used a typical greenery exposure measure from remote sensing data. The normalized difference vegetation index (NDVI) was derived from the Landsat 8 Operational Land Imager and the Thermal Infrared Sensor at a spatial resolution of 30 meters. Data were obtained for the year 2016 from the USGS EarthExplorer (https://earthexplorer.usgs.gov/). To avoid distortions in the NDVI, cloud-free images in the greenest season (i.e., June–August) were used. Based on the land surface reflectance of the visible red channel (RED) and the near-infrared channel (NIR), the NDVI is calculated as follows: (NIR-RED) / (NIR+RED) [29,47]. The index ranges from −1 to 1; more positive values refer to greener vegetation, and negative values refer to non-biomass (e.g., water). In order to avoid distortions due to negative values when averaging NDVI values per neighborhood, negative values were disregarded [15,30].

*Potential mediators*

Four potential mediators were obtained from the questionnaire as suggested in the literature [14]. First, respondents' physical activity was quantified by their weekly physical exercise time in hours [19]. Second, self-reported stress [24] was measured by one question. The variable was coded into five categories. Third, the perceived air quality and noise [17,48] in the neighborhood were assessed on the basis of two 5-level Likert items. For the final index, both items were averaged; higher values indicate more environmental satisfaction. Fourth, social cohesion [24,26] was measured by six 5-level items. Cronbach's alpha for the items was 0.82. We determined the respondents' mean scores on these items. Each question is given in Table A2 in the supplementary materials.

*Covariates*

Widely employed covariates referring to people's demographic and socioeconomic circumstances as well as their lifestyles were included a priori. Our controls were gender (male; female), age (in years), education attainment (primary school or below; high school; college or above), marital status (married; single, divorced, or widowed), hukou (i.e., the household registration system in China) status (yes; no), household wealth expressed through four income categories (in Chinese yuan, CNY), household size (number of people), length of occupancy (in years), functional ability (restricted; not restricted), medical insurance status (yes; no), smoker (yes; no), and drinker (yes; no).

**2.3 Statistical analyses**

Summary statistics were employed to describe the study population and to assess the central features of each variable. A non-parametric Spearman correlation was used to measure the association between streetscape greenery and NDVI on a neighborhood-level.

To examine whether the association between greenery and mental wellbeing could be partially or fully explained by the mediators—namely physical activity, stress, air quality and noise, and social cohesion—we conducted a multi-step mediation analyses [49]. Mediation analysis decomposes the effect of greenery on mental wellbeing into a direct and a mediation (indirect) component (i.e., the total effect minus the direct effect) as well as a total effect. The percentage of mediation was determined by dividing the mediation by the total effect.

The following steps were conducted. First, we identified the direct association between neighborhood-based greenery and respondents' WHO-5 scores. Model 1a included streetscape greenery, Model 1b included the NDVI, Model 1c included both greenery measures together, and Model 1d included a streetscape greenery-NDVI interaction term. We fitted fully adjusted multilevel linear regression models [50] considering WHO-5 as the continuous outcome. In order to consider correlations that arise because respondents are nested in neighborhoods with similar exposures, a random intercept was included. Model performance was assessed by the Akaike information criterion (AIC); a smaller AIC score refers to a better goodness-of-fit [51]. Second, in Model 2a we regressed each mediator on streetscape greenery and in Model 2b on the NDVI. These models were fully adjusted. Third, the Models 3a, c, e, g, and i determined the association between streetscape greenery and WHO-5 scores including the significant mediators one at a time and simultaneously from the second step. The Models 3b, d, f, h, and j assessed the NDVI-WHO-5 association and the mediators separately and together. Figure 1 summarizes our conceptual model.

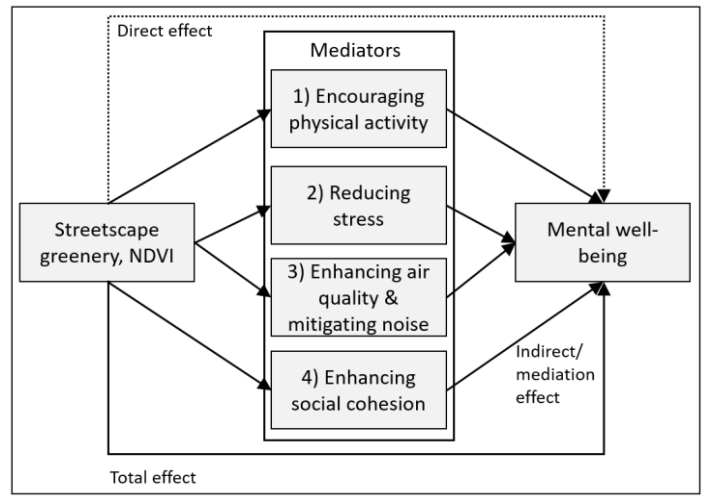

Figure 1. Greenery–mental wellbeing mediation model. All pathways were adjusted for multiple covariates.

Partial mediation occurs when the greenery effect from step 3 is reduced compared to step 1. Complete mediation occurs when the significance of the effect disappears entirely. To test the significance of the mediation effect, Sobel tests were conducted [52]. The test statistic assesses whether including a mediator in the regression together with streetscape greenery or NDVI reduces the greenery effect while the mediator remains significantly correlated. Statistical analyses were carried out in STATA 15.1.

## 3 Results

### 3.1 Descriptions of the study population and the greenery measures

The characteristics of the study population are summarized in Table 1. A comparison between the characteristics of our sample and the general population (>18 years old) in the inner city of Guangzhou showed good agreement (Table A1). The 1,029 participants had a mean age of 41 years; half of them (50.146%) were male (standard deviation (SD) ±13.576). On average, the WHO-5 score was 12.1, with a SD of ±3.7. The average NDVI score was 0.131 (SD±0.113), while the proportion of streetscape greenery was 0.267 (SD±0.060). These differences across both greenery measures are also apparent in Figure A1 (supplementary materials). Respondents reported a mean physical activity level of 4 hours per week; the SD (±4.2) was large. The subjects seldom experienced stress (mean=1.7, SD±0.808). A mean of 3.1 (SD±1.050) refers to a neutral perception of air quality and noise within the neighborhood; the rating for social cohesion was also neutral (mean=2.948, SD±0.582). The Spearman correlation (0.045) was positive between both greenery measures, but did not reach statistical significance ($p$=0.789).

Table 1. Summary statistics of the study population.

| Variables | Mean (SD) | 25% quantile | 75% quantile | Proportion |
|---|---|---|---|---|
| *Response* | | | | |
| WHO-5 score | 12.081 (3.706) | 10.000 | 14.000 | |
| *Greenery exposure measures* | | | | |
| NDVI score | 0.131 (0.113) | 0.074 | 0.117 | |
| Streetscape greenery (%) | 0.267 (0.060) | 0.209 | 0.323 | |
| *Mediators* | | | | |
| Physical activity (hours/week) | 4.024 (4.182) | 1.000 | 6.000 | |
| Social cohesion | 2.948 (0.582) | 2.500 | 3.333 | |
| Air quality and noise | 3.058 (1.050) | 2.000 | 4.000 | |
| Stress | 1.798 (0.808) | 1.000 | 2.000 | |
| *Covariates* | | | | |
| Gender (%): Male | | | | 50.146 |
| Female | | | | 49.854 |
| Age | 41.185 (13.576) | | | |
| Marital status (%): Single, divorced, or widowed | | | | 21.672 |
| Married | | | | 78.328 |
| Hukou status (%): Local hukou | | | | 77.745 |
| Non-local hukou | | | | 22.255 |
| Education attainment (%): Primary school or below | | | | 2.527 |
| High school | | | | 50.048 |
| College or above | | | | 47.425 |
| Household wealth (%):<10,000 CNY | | | | 7.192 |
| 10,000–<20,000 CNY | | | | 70.651 |
| 20,000–<40,000 CNY | | | | 15.258 |
| ≥40,000 CNY | | | | 6.899 |

| | | |
|---|---|---|
| Household size (persons) | 3.282 (0.852) | |
| Length of occupancy (years) | 13.206 (11.021) | |
| Functional ability (%): restricted | | 3.984 |
|   Not restricted | | 96.016 |
| Medical insurance (%): Yes | | 97.085 |
|   No | | 2.915 |
| Smoker (%): Yes | | 26.433 |
|   Non-smoker | | 73.567 |
| Drinker (%): Yes | | 41.108 |
|   No | | 58.892 |

*3.2 Correlations between greenery and WHO-5*

With a neighborhood-level intra-class correlation of 0.385 of the null model (i.e., one without any variable), the application of multilevel models was justifiable. Table 2 displays the results of the first step of our mediation analysis (i.e., the total effects of greenery on WHO-5) (Model 1a and b). With an AIC difference of 4.922, Model 1b with NDVI had a better goodness-of-fit than Model 1a with streetscape greenery. Independent of the exposure measures, we consistently observe that neighborhood greenery is positively associated with WHO-5 scores (i.e., the more, the better). Model 1c assessed the joint effect of both greenery measures. Both greenery measures were significant. The interaction effect between the greenery measures turned out to be insignificant (Model 1d). Concerning the covariates, no notable differences occurred between the models. Local hukou, household wealth, length of occupancy, medical insurance, and smoking were not significant.

Table 2. Regression results showing the total effects of greenery on WHO-5.

| | Model 1a | Model 1b | Model 1c | Model 1d |
|---|---|---|---|---|
| | Coef. (SE) | Coef. (SE) | Coef. (SE) | Coef. (SE) |
| Streetscape greenery | 3.768*** (1.153) | | 3.656*** (1.005) | 4.245*** (1.303) |
| NDVI | | 9.960** (4.281) | 10.011** (4.284) | 10.121** (4.585) |
| Male (Ref.: Female) | 0.785*** (0.294) | 0.784*** (0.294) | 0.784*** (0.293) | 0.773*** (0.293) |
| Age | -0.005 (0.010) | -0.006 (0.010) | -0.005 (0.010) | -0.006 (0.010) |
| Married (Ref.: Single, divorced, and widowed) | -0.603** (0.277) | -0.598** (0.277) | -0.598** (0.276) | -0.580** (0.276) |
| Local hukou (Ref.: Non-local hukou) | -0.244 (0.233) | -0.244 (0.233) | -0.245 (0.232) | -0.238 (0.232) |
| Education (Ref.: Primary school or below) | | | | |
|   High school | 0.294 (0.632) | 0.296 (0.632) | 0.294 (0.632) | 0.301 (0.632) |
|     College and above | 1.171* (0.675) | 1.165* (0.675) | 1.163* (0.674) | 1.156* (0.674) |
| Household wealth (Ref.: <10,000 CNY) | | | | |
|     10,000-<20,000 CNY | -0.087 (0.387) | -0.074 (0.387) | -0.074 (0.387) | -0.068 (0.387) |
|     20,000-<40,000 CNY | 0.131 (0.459) | 0.157 (0.459) | 0.157 (0.458) | 0.167 (0.458) |
|     >40,000 CNY | 0.172 (0.513) | 0.157 (0.513) | 0.158 (0.512) | 0.158 (0.512) |
| Household size | -0.173 (0.123) | -0.233* (0.123) | -0.232* (0.122) | -0.221* (0.122) |
| Occupancy of residence | 0.002 (0.011) | 0.003 (0.011) | 0.003 (0.010) | 0.003 (0.010) |
| Functional ability (Ref.: Not restricted) | -1.545*** (0.500) | -1.535*** (0.500) | -1.534*** (0.499) | -1.551*** (0.499) |
| Medical insurance (Ref.: None) | -0.032 (0.575) | -0.031 (0.574) | -0.028 (0.574) | -0.035 (0.574) |
| Drinking (Ref.: No) | -0.737** (0.310) | -0.742** (0.310) | -0.742** (0.310) | -0.736** (0.310) |
| Smoking (Ref.: No) | -0.319 (0.250) | -0.311 (0.250) | -0.310 (0.250) | -0.310 (0.250) |
| Streetscape greenery × NDVI | | | | 0.274 (0.723) |
| Constant | 13.547*** (2.306) | 12.572*** (1.123) | 12.146*** (2.224) | 11.305*** (1.948) |
| Variance (Neighborhood level) | 5.266** | 4.388** | 4.392** | 4.383** |
| Variance (Individual level) | 7.775** | 7.781** | 7.781** | 7.790** |
| Log likelihood | -2,522.239 | -2,519.781 | -2,529.756 | -2,523.937 |
| AIC | 5,082.485 | 5,077.563 | 5,089.514 | 5,079.874 |

Significance levels: * $p<0.10$, ** $p<0.05$, *** $p<0.01$. SE=standard error.

### *3.3 Correlations between greenery and possible mediators*

Table 3 shows the results of regressing greenery on the mediators. In Models 2a and 2b, greenery was positively correlated with physical activity. More streetscape greenery was also related to reduced perceived psychological stress; however, for NDVI the association was counterintuitive but insignificant. We observed that streetscape greenery was positively correlated with perceived air quality and noise at a 5% level; it was insignificant for NDVI. Respondents' social cohesion score was significantly positively correlated with streetscape greenery and NDVI.

Table 3. Results of regressing greenery on the mediators.

| | | Physical activity | Stress | Air quality and noise | Social cohesion |
|---|---|---|---|---|---|
| Streetscape greenery | Coef. | 24.397*** | -3.082** | 6.086** | 1.050** |
| (Model 2a) | (SE) | (8.766) | (1.407) | (2.471) | (0.430) |
| NDVI | Coef. | 10.028** | 1.623 | 0.116 | 0.144** |
| (Model 2b) | (SE) | (4.325) | (0.935) | (1.537) | (0.065) |

Significance levels: * $p<0.10$, ** $p<0.05$, *** $p<0.01$. SE=standard error. Models were fully adjusted.

### *3.4 Correlations between mediators, greenery, and WHO-5*

Table 4 summarizes the results concerning whether the greenery–WHO-5 associations were mediated. Only the significant mediators from Table 3 were included. Even after considering these mediators, streetscape greenery and NDVI remained significantly and positively related to WHO-5 scores. There seems to be significant partial mediations of physical activity on streetscape greenery and NDVI–mental wellbeing correlations (Models 3a and 3b) as indicated by the Sobel test ($Z_{SC}$=2.412, $p$=0.016; $Z_{NDVI}$=2.077, $p$=0.038). Respondents' WHO-5 scores decreased with increasing perceived stress, which was a partial mediator for streetscape greenery (Model 3c) but not for NDVI. The mediation was significant ($Z_{SC}$=2.050, $p$=0.040). The Sobel test was also significant for perceived air quality and noise ($Z_{SC}$=2.050, $p$=0.040) (Model 3e). Across both greenery measures (Models 3g and 3h) social cohesion was positively correlated with people's WHO-5 scores and a significant mediator ($Z_{SC}$=2.388, $p$=0.017; $Z_{NDVI}$=2.183, $p$=0.029). Considering the significant mediators in combination (Model 3i and 3j) did not alter our findings. Only the coefficients reduced slightly compared to the models with a single mediator one at a time (Table 4).

Table 4. Results of the greenery–mental wellbeing relation: The mediating effect of physical activity, stress, air quality and noise, and social cohesion.

|  | Model 3a | | Model 3c | | Model 3e | | Model 3g | | Model 3i | |
|---|---|---|---|---|---|---|---|---|---|---|
|  | Coef. | (SE) | Coef. | (SE) | Coef. | (SE) | Coef. | (SE) | Coef. | (SE) |
| Streetscape greenery | 3.210*** | -1.153 | 3.155*** | -1.154 | 2.998*** | -1.148 | 3.398*** | -1.158 | 2.226** | -1.112 |
| Physical activity | 0.121*** | (0.025) | - | | - | | - | | 0.082*** | (0.024) |
| Stress | - | | -0.775*** | (0.133) | - | | - | | -0.582*** | (0.124) |
| Air quality and noise | - | | - | | 0.376*** | (0.131) | - | | 0.303*** | (0.124) |
| Social cohesion | - | | - | | - | | 2.100*** | (0.164) | 1.921*** | (0.167) |
| AIC | 5,062 | | 5,042 | | 5,076 | | 4,931 | | 4,897 | |
|  | Model 3b | | Model 3d | | Model 3f | | Model 3h | | Model 3j | |
|  | Coef. | (SE) | Coef. | (SE) | Coef. | (SE) | Coef. | (SE) | Coef. | (SE) |
| NDVI | 8.792** | -4.202 | - | | - | | 9.760*** | -3.730 | 8.602** | -3.694 |
| Physical activity | 0.117*** | (0.025) | - | | - | | - | | 0.077*** | (0.024) |
| Stress | - | | - | | - | | - | | - | |

| | | | | | | | |
|---|---|---|---|---|---|---|---|
| Air quality and noise | - | - | - | - | - | - | - |
| Social cohesion | - | - | - | 2.106*** | (0.163) | 2.034*** | (0.164) |
| AIC | 5,058 | - | - | 4,925 | | 4,917 | |

Coef.=coefficient; SE=standard error. The outcome variable were the WHO-5 scores. Models were adjusted for individual covariates. Significance levels: * $p<0.10$, ** $p<0.05$, *** $p<0.01$.

Finally, Table 5 summarizes the direct, mediating, and total effects of the mediation analyses. For streetscape greenery, physical activity accounted for the largest proportion (47.906%) of mediation, and social cohesion for the smallest (39.353%). Stress (43.081%) and air quality and noise (43.284%) were well balanced. Pronounced reductions in the proportion of mediation were observable for NDVI. Here, the proportion mediated by physical activity was 11.771%; for social cohesion it was 3.011%. In the multiple mediation models, the mediators combined accounted for 62.162% (streetscape greenery) and 22.196% (NDVI).

Table 5. Direct, indirect, and total effects of the mediation analyses

| Streetscape greenery | Direct effect | Indirect effect | Total effect | % mediation effect |
|---|---|---|---|---|
| Physical activity | 3.210 | 2.952 | 6.162 | 47.906 |
| Stress | 3.155 | 2.388 | 5.543 | 43.081 |
| Air quality and noise | 2.998 | 2.288 | 5.286 | 43.284 |
| Social cohesion | 3.398 | 2.205 | 5.603 | 39.353 |
| Multiple mediators | 2.226 | 3.657 | 5.883 | 62.162 |
| NDVI | Direct effect | Mediating effect | Total effect | % mediation effect |
| Physical activity | 8.792 | 1.173 | 9.965 | 11.771 |
| Social cohesion | 9.760 | 0.303 | 10.063 | 3.011 |
| Multiple mediators | 8.602 | 2.454 | 11.056 | 22.196 |

**4 Discussion**

This study examined how four mediators—namely physical activity, stress, air quality and noise, and social cohesion, both individually and in combination—affect the correlation between greenery and mental wellbeing. This study is unique in that it incorporated the underlying mechanisms for greenery that were automatically audited on site and captured through remote sensing.

### 4.1 Main findings

Our finding that more greenery correlates positively with mental health and wellbeing corresponds with findings reported elsewhere [6,25,27]. [13] observed that while NDVI was positively associated with wellbeing for two Australian cities, this was not the case for two cities in New Zealand. The latter also corresponds to a Dutch study on psychotic disorders [11]. Regardless of these mixed findings, a systematic review found a stronger tendency that exposure to more greenery seems to be beneficial for mental health and wellbeing [2].

Although the application of street view data is gaining momentum [32,33,36], the evidence base for streetscape greenery is weak. We are aware of only a few studies that assessed streetscape greenery on mental health [24,34,35,53], and only one employed a similar deep learning approach [31]. Our finding that both streetscape greenery and NDVI support mental wellbeing among adults is novel. Most importantly, both greenery measures reached statistical significance in the regressions even though they correlated only weakly with each other, as others reported [31,54]. Moreover, assessing joint effect of both measures and their interaction provide further evidence that streetscape greenery and NDVI represent different aspects of the natural environment. This many also be an explanation why the coefficients across the greenery measures differ markedly; however, this conclusion still remains speculative. That exposure to streetscape greenery may guard against mental disorders corroborates a study among elderly people in Beijing, China [31] and one in Ottawa, Canada [34]. It seems that small-sized natural elements (e.g., trees) and/or vertical natural elements (e.g., green walls) are beneficial to resident' health [24,35,55]. In the Beijing study, however, NDVI did not reach statistical significance [31].

To shed light on the still hypothetical mechanisms, mediation analyses were conducted. We found that the streetscape greenery–mental wellbeing correlation was mediated by physical activity, stress, air quality and noise, and social cohesion, while only physical activity and social cohesion appeared to be a mediator between NDVI and mental wellbeing. In line with our hypothesis, perceived streetscape greenery, often small in size, makes a significant contribution to mental wellbeing, by buffering against stress and mitigating environmental pollutants (i.e., air pollution and noise), which seems essential in highly urbanized inner cities where larger green spaces are scarce. Although not exactly comparable with prior studies due to differences in research designs, that approximately one fifth of the NDVI correlations explained by the mediators mirrors a European study [19]. For streetscape greenery, the proportion (two thirds) was considerable higher. In the context of the available evidence, the underlying reasons for such differences deserve more attention, but our results support the hypothesis that streetscape and remote sensing-based assessments may indeed capture diverse greenery aspects.

As we are unaware of another mediation study like ours, the following discussion is not specifically tailored to streetscape greenery. In line with our results, two Dutch studies [24,25] report that pronounced social cohesion contributes to the mental health benefits of greenery, which contradicts findings from adults in Catalonia, Spain [27]. Consistent across the greenery measures, we found that physical activity may be among the mediators. Greenery may motivate people to undertake physical activities (e.g., walking; [44]), which supports wellbeing [56]. However, due to a lack of correlations between physical activity, social support, and greenery, others [24,27] questioned such a possible pathway. In contrast, an Australian study found that physical activity in the form of recreational walking explained parts of the association between greenery and mental health [26]. In our models, perceived stress was a significant mediator only for streetscape greenery, supporting the notion that the health-supportive effect of greenery may be via stress reduction, as also pointed out in other studies [14,27]. We showed that perceived air quality and noise may be among the mechanisms whereby streetscape greenery promotes mental health and wellbeing. That the mediator was not of relevance for NDVI is counterintuitive, as vegetation absorbs air pollutants and noise [21–23]. The differences in our mediation analyses across both greenery measures verify our aforementioned findings that different operating mechanisms could be at play.

**4.2 Strengths and limitations**

Our study has multiple strengths and limitations. It is the first population-based study we are aware of that conducted mediation analysis with street view greenery data in the context of mental health. A methodological strength is the way in which we modeled greenery, namely by coupling cutting-edge deep learning with street view data. Neither streetscape greenery nor NDVI is affected by people's self-reporting and subjective perception [57], though it raises questions about the actual greenery use [58]. Both measures also do not convey information about the quality of greenery [58]. Where earlier studies [7,12] focused on direct correlations between greenery and mental health, we centered this analysis, like only a few others [17,19,20], on potential mediators affecting this association. Related to this, testing the four suggested mechanisms simultaneously [14] is another strength; some previous studies tested only two [27]. Unlike concentrating on Western countries, a vital aspect was the selection of a rapidly urbanizing Chinese metropolis. However, transferring and generalizing our results to other areas requires verification.

Privacy issues prevented us from assessing greenery in the immediate vicinity of people's homes [4], though it is likely that greenery varies within neighborhoods. Moreover, since no information on people's activity locations and daily travel was available (e.g., GPS tracks; [33]), our exposure assessment may be biased [59]. Future studies are urged to address this issue. When working with street view images, it is unavoidable that results may be determined by the availability of images. Such images are usually collected from moving vehicles; greenery apart from roads is usually not included. Many street view images are required to approximate streetscape greenery, which increases the computational burden for large nationwide studies. Our street view greenery assessment did not consider that distances between the street center lines and the objects (e.g., facades) could vary. This may have affected the amount of greenery per image. Like elsewhere [17,23], some variables (i.e., noise) were incorporated through people's subjective experiences, which are likely to deviate from objective measurements [57]. To facilitate comparability with previous studies [24], we employed mediation analysis after Baron and Kenny [49]. We acknowledge that this approach is not without criticism [60] and more flexible methodologies may serve as alternative [61]. Despite our efforts in adjusting for key personal and lifestyle factors, some confounders were likely to be missing. Due to the lack of information on people's attitudes toward and motives for selecting a residential neighborhood, self-selection remains an issue [62]. Finally, the cross-sectional nature precluded us from making causal statements and reverse-causality cannot be ruled out.

## 5 Conclusion

This cross-sectional study assessed greenery–mental health and wellbeing pathways. While streetscape greenery and remotely sensed greenery were insignificantly correlated, our regression results coherently suggest that pronounced exposure to greenery, independent of the measure, is related to gains in people's mental wellbeing. Mediation analyses revealed striking differences between the two exposure metrics in pathways underlying the effect of greenery exposure on mental wellbeing. Physical activity, stress, air quality and noise, and social cohesion partially mediated the relation between streetscape greenery and mental wellbeing; only physical activity and social cohesion served as partial mediators of the health benefits of NDVI. The explanatory power of the mediators was substantial, namely 62% for streetscape greenery and 22% for NDVI.

Taken together, our findings provide evidence that both greenery measures signify different aspects of natural environments. We infer that both measures have different operating mechanisms. Environmental health managers are advised to conduct urban greening interventions in public spaces and to preserve smaller-scale and vertical greenery, as they may represent another way to gain health benefits on a population-level by manipulating the mediators. Further research on streetscape greenery is urged to replicate our findings.

**Acknowledgements**

We thank the anonymous reviewers for their valuable comments on an earlier draft of this paper.

**Funding**

MH was funded by the European Research Council (ERC) under the European Union's Horizon 2020 research and innovation program (grant agreement No 714993). This work was supported by the National Natural Science Foundation of China (grant numbers 41871140, 41801306, and 51678577) and by the Innovative Research and Development Team Introduction Program of Guangdong Province awarded to YL.

**Author contributions**

RW and MH developed the research idea. YL, JZ, PL, and YY collected and organized the street view data. YY developed the deep learning model. RW carried out the statistical analysis. MH wrote the manuscript with some input from RW. MH, RW, and YL revised the manuscript. All authors approved the final manuscript.

**Declaration of interest**

None.

**Conflict of interest**

The authors have no conflict of interest to declare.

Table S1. Summary statistics of our sample compared to people in the inner city of Guangzhou (census data 2010).

| Variables | Proportion | |
| --- | --- | --- |
| | Census | Samples |
| Gender (%) | | |
|   Male | 51 | 50 |
|   Female | 49 | 50 |
| Education (%) | | |
|   College or above | 30 | 47 |
|   High school | 60 | 50 |
|   Primary school or below | 10 | 3 |
| Marital status (%) | | |
|   Married | 81 | 79 |
|   Single, divorced, or widowed | 19 | 21 |
| Age (%) | | |
|   20-29 | 31 | 25 |
|   30-39 | 24 | 26 |
|   40-49 | 21 | 26 |
|   50-59 | 10 | 9 |
|   60-69 | 6 | 9 |
|   70-79 | 4 | 5 |

Table A2. Survey questions (originally in Chinese).

| Variable | Question | Items |
|---|---|---|
| WHO-5 score | Indicate for each of the 5 statements which is closest to how you have been feeling over the past 2 weeks. | 0=At no time |
| | | 1=Some of the time |
| | 1. I have felt cheerful and in good spirits | 2=Less than half the time |
| | 2. I have felt calm and relaxed | 3=More than half the time |
| | 3. I have felt active and vigorous | 4=Most of the time |
| | 4. I woke up feeling fresh and rested | 5=All of the time |
| | 5. My daily life has been filled with things that interest me | |
| Physical activity | What is your last week's total exercise time (hours/week)? | |
| Social cohesion | Do you agree with the following statement about the neighborhood? | 1=Strongly disagree |
| | 1. Neighbors often drop in on each other | 2=Disagree |
| | 2. People around here are willing to help their neighbors | 3=Neutral |
| | 3. People in this neighborhood always share health information with each other | 4=Agree |
| | 4. People in this neighborhood greet each other when they meet | 5=Strongly agree |
| | 5. People in this neighborhood can be trusted | |
| | 6. People can deal with problems in the neighborhood together | |
| Air quality and noise | Please rate your satisfaction with the following factors in your neighborhood | 1=Strongly unsatisfied |
| | 1. Air quality | 2=Unsatisfied |
| | 2. Noise | 3=Neutral |
| | | 4=Satisfied |
| | | 5=Strongly satisfied |
| Stress | How often did any emotional problems (i.e., feeling depressed or anxious) affect your work or daily activities in the past month? | 1=Never |
| | | 2=Seldom |
| | | 3=Sometimes |
| | | 4=Often |
| | | 5=Always |

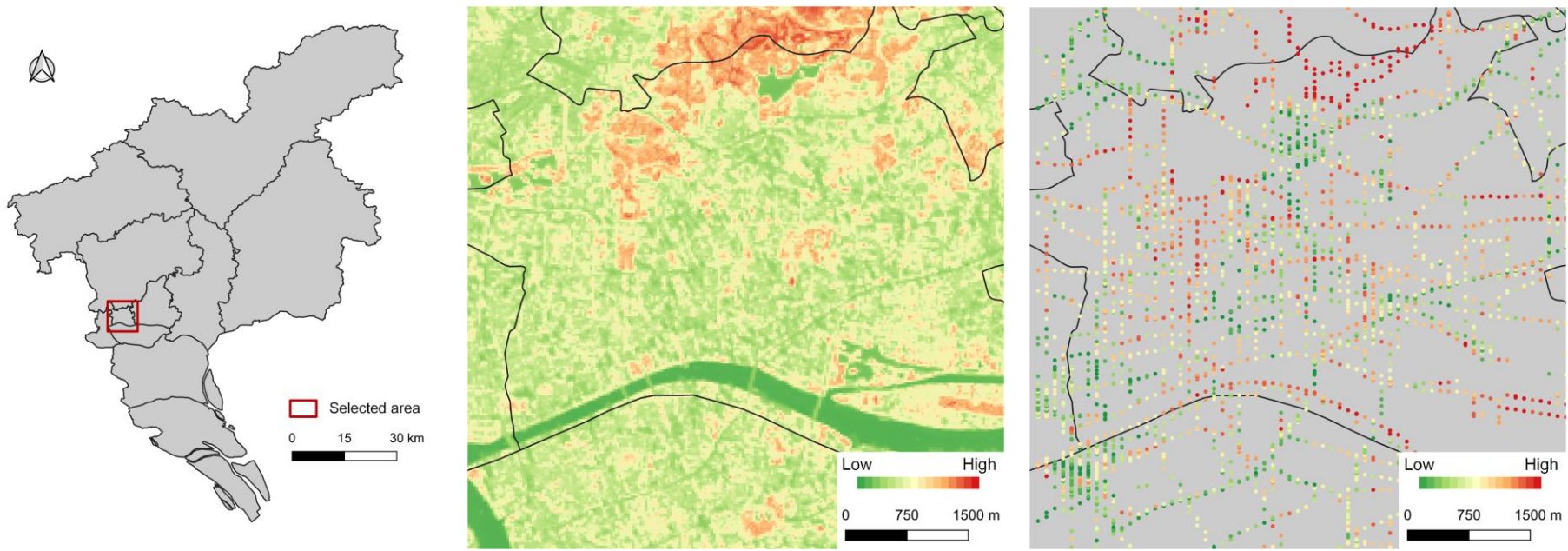

Figure A1. A comparison of NDVI (middle panel) and streetscape greenery (right panel) for a selected area in Guangzhou.